# OBSERVATION OF FALSE SPHERICAL MICROMETEORITES

Attilio Anselmo

Stoppani S.P.A., Cogoleto (GE) 16016 Italy (e-mail: ansatt@libero.it)

The work describes the results of the study of the spherical particles that can be found in the environment and that were often considered as micrometeorites. The results have demonstrated that in the most of cases these spherical particles are the results of the human activity.

Micrometeorites, found in the South Pole water well [1] and in the ice of Greenland [2] has attracted an attention of scientists as they can serve for the determination of the earth atmosphere evolution. However, during the last decade, several works appeared, suggesting the simple way of the collection of micrometeorites in the human environment, in particular, on the roofs of houses (for example, see [3, 4]). As it is written in these works, these meteorites must have a spherical shape, their dimensions must be in the range from tens of microns till millimeter, and they must have magnetic properties. As it is also indicated, there is, however, one source of possible artifacts – soldering. These publications were very interesting for me and I have decided to study these objects. In fact, when I have collected the powder on the roof and have analyze it with a microscope, among other different objects, I have founded spherical particles, that can be magnetized. Image, showing the general picture of what can be collected from the roof, is shown in Fig. 1a. Spherical particles (sometimes, hollow) of the same sizes, mentioned above, are clearly visible (see zoomed part of the Fig. 1a). However, there are two other types of object frequently present in the samples collected on the roofs: glass-like fibers (Fig. 1b) and yellow-brown colored granules with clearly visible sharp corners (Fig. 1c).

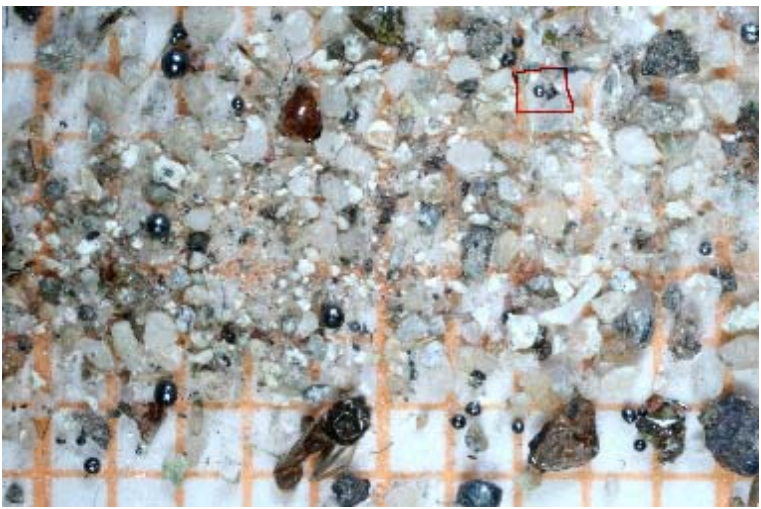
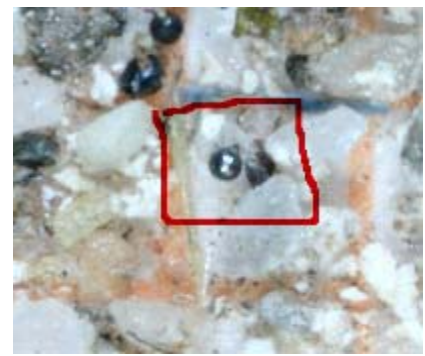

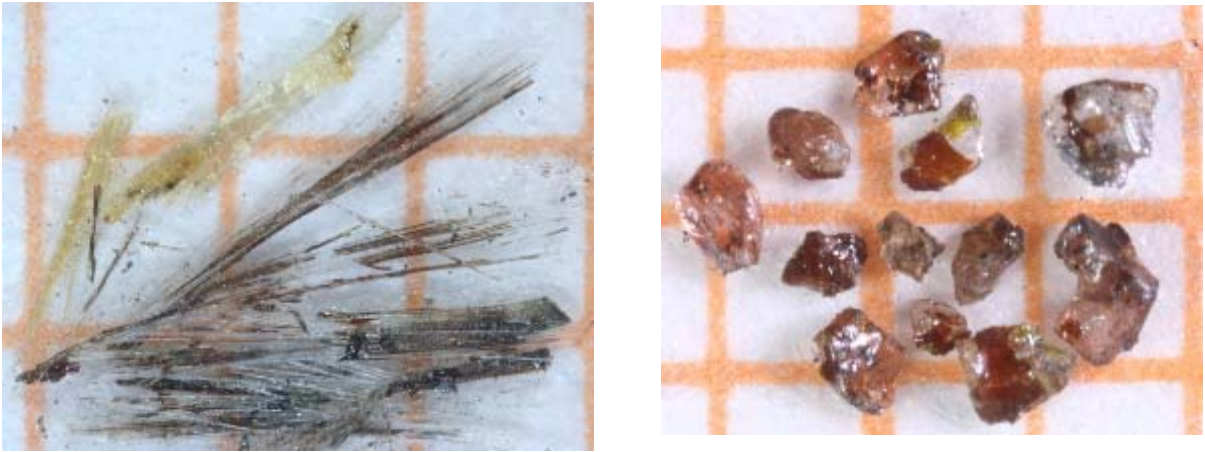

Fig. 1. Images of the samples collected on the roofs. (a) Image representing row sample (zoomed area demonstrates the presence of spherical particles); (b) Image of fibers, separated from row samples; (c) Image of granules, separated from row samples. Grid unit corresponds to 1 mm.

Even if the characteristics of the collected spherical particles were absolutely the same, attributed to the micrometeorites, the further investigations, directed to the statistical study of their distribution in the environment, have revealed several observation, listed below, making very questionable the extra-terrestrial origin of these spherical particles:
1. The distribution of these spherical particles is very inhomogeneous: their concentration is very high in areas where people live (and even higher in the industrial zones) and they are practically absent in areas located fare from the urbane areas (few present are very small in sizes).
2. In areas closed to the zones of human activity, fibers and granules, similar to those shown in Fig. 1b and Fig. 1c, can be frequently found.
3. There is inhomogeneous distribution of these spherical particles also in height: they can be frequently found on roofs of houses of 3-4 floors height, and, even in industrial zones, very few spherical particles with reduced diameter can be found on roofs of buildings higher than 20 floors.

The listed observations corroborate strongly the terrestrial origin of these particles.

The other observation against the cosmic origin of these "micrometeorites" was performed in November 1998. In this period (17-18 November), Leonids have arrived to the earth, and the increased number of micrometeorites was expected [5]. Therefore, November 18 1998, I have asked a lot of persons living in different zones of Italy to collect the dust from their roofs and to send them to me. More than 20 samples, collected in different zones of center and north of Italy were analyzed. However, no one sample had revealed the increased concentration of these spherical particles.

Considering all the presented data, the most probable logical conclusion was that these spherical particles are due to the human industrial activity. As it was mentioned, the soldering must be the only source of the "false micrometeorites" origin. Thus, I have decided to investigate this problem.

Next samples were taken in the mechanical workshop equipped also with soldering machines. The image of the sample collected in the workshop is shown in Fig. 2.

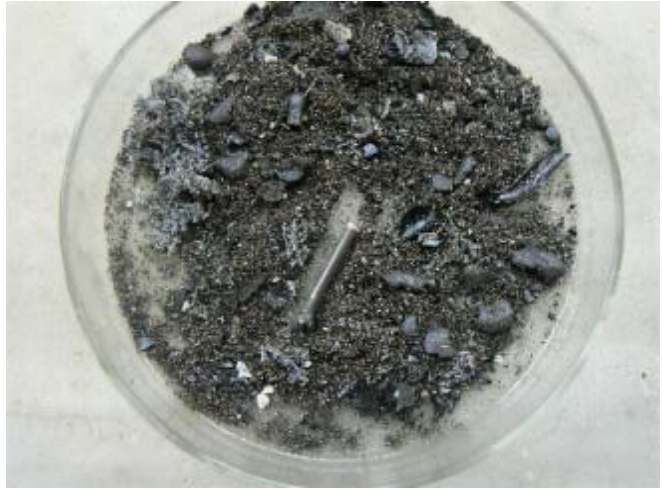

Fig. 2. Image of row sample collected in the mechanical workshop.

The analysis have revealed the presence of spherical particles and also the presence of significant amount of fibers and granules, similar to those, shown in Fig 1 b and Fig. 1 c. The search of the source of these last object has brought me to the grinding wheel also present in the mechanical workshop. In fact, the structure of this object contains both fibers and granules. The image of the cut of the grinding wheel is shown in Fig. 3.

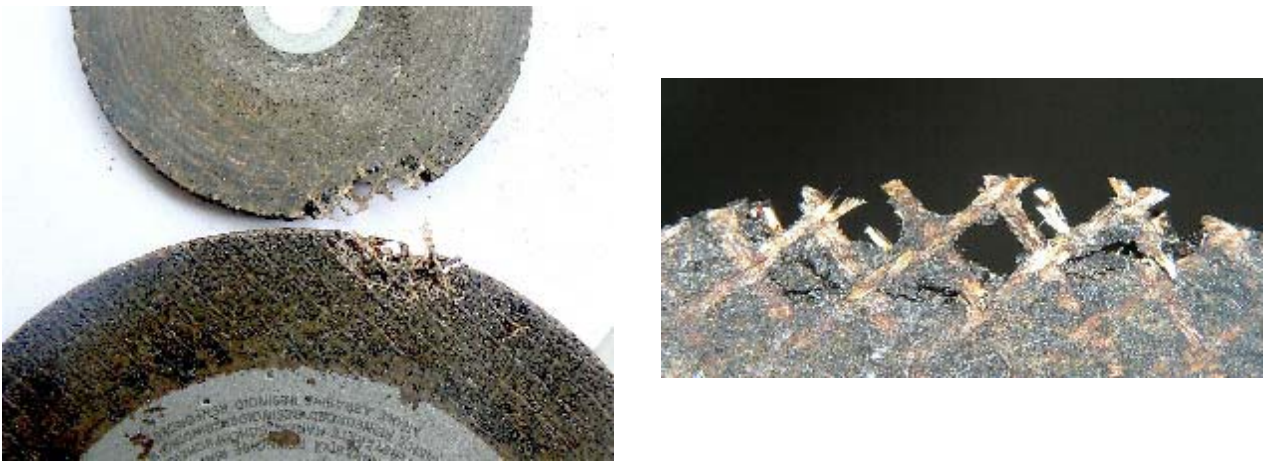

Fig. 3. General view and zoomed area of the grinding wheel.

Thus, the logic question appeared: can it be a grinding wheel treatment of the metals a source of the observed spherical particles? As everybody knows, the sparks produced during this treatment (Fig. 4) are small pieces of melted metal, at a very high temperature, reacting with atmospheric oxygen (Fig. 5). Therefore, if steel was treated, we can expect the formation of iron oxide particles [6, 7], that can have magnetic properties and will not be further affected by corrosion. Their spherical shape is due to the melting and corrosion resistance is due to the fact that are already oxidized.

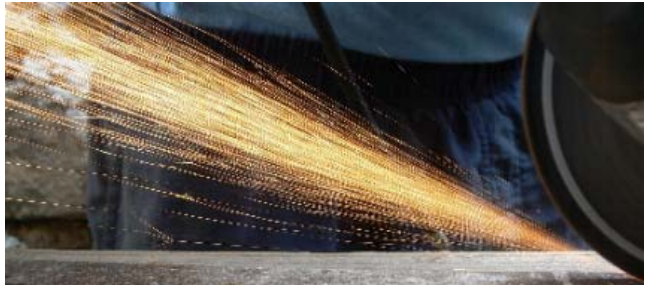

Fig. 4. Sparks produced during grinding wheel treatment.

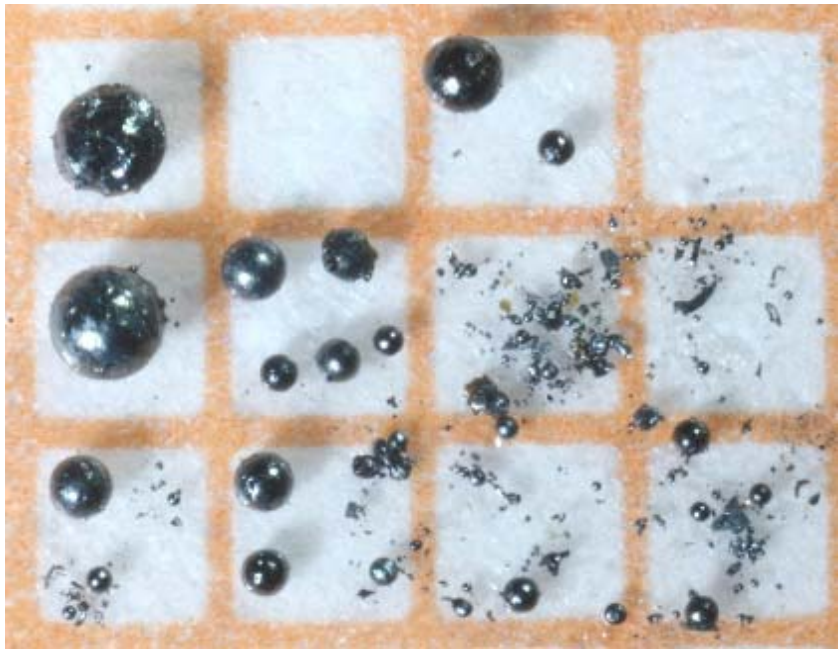

Fig. 5. Spherical particles produced as a result of grinding wheel treatment. Grid unit corresponds to 1 mm.

Collected cooled sparks resulted from the treatment of different materials were used as next samples. Acquired images are shown in Fig. 6. As one can see, the sizes and shape of the spherical particles are strongly dependent on the type of the material, especially, on its hardness. The observation does not seem strange. In fact, as the increase effort must be applied to cut the pieces of more harder material and, as a result, the initial temperature of the produced sparks will be higher. Therefore, usually we can see increased concentration of the formed spherical particles when working with hard materials, as the applied effort must be higher. However, its concentration will be the same if the same effort is applied for the softer material treatment. In fact, the melting temperatures of all these materials are similar. Thus, the applied effort is a key parameter of the detached pieces temperature and, therefore, determines how many particles will be melted and will form spherical particles. The fragility of the material is also important. For example, samples of cast iron and Widia have revealed very low concentration of the spherical particles (very few sparks were also produced during the treatment of these materials).

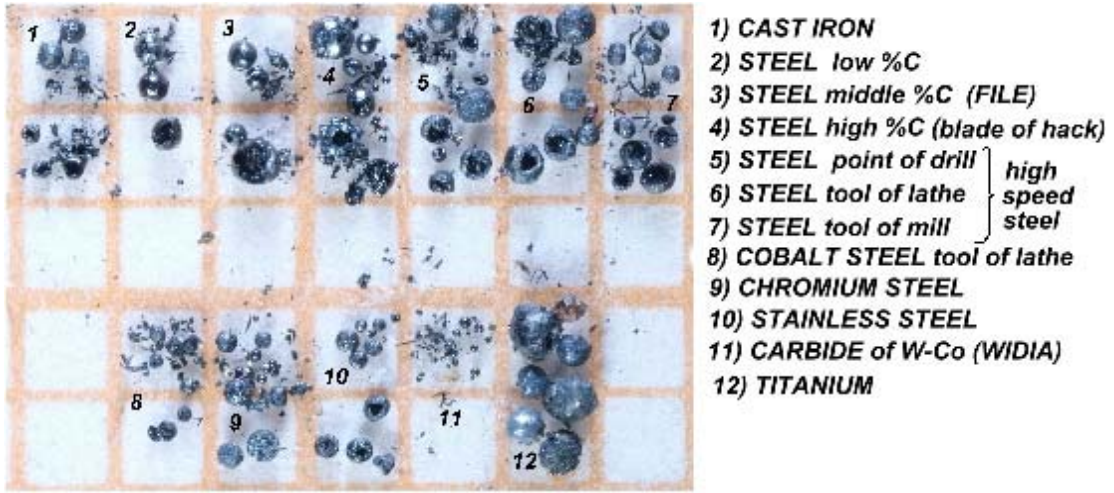

Fig. 6. Spherical particles produced as a result of grinding wheel treatment of different types of samples (see the legend). Grid unit corresponds to 1 mm.

The formation of hollow particles can be explained by the explosive thermal expansion of $CO_2$, formed as a result of the interaction of the internal carbon of cast iron or steal with oxygen at high temperature.

The presented results have demonstrated that the grinding wheel treatment can be considered as the source of spherical particles that were often considered as micrometeorites. The spherical particles usually considered as micrometeorites are similar to those (2, 3 and 4) in Fig. 5. In fact, these spherical particles were obtained from the mechanical treatment that is often performed in the open areas. All the other samples (1, 5-12) are the results of the treatment that was performed in the closed mechanical workshops, and, therefore, the probability to find them is much less.

When the conclusion was performed, analysis were carried out also in closed areas. It was established that similar spherical particles can be found in the dust within closed areas (rooms).

There is still one question to answer. Why we can find such spherical particles rather far from the industrial activity zones but close to the people living areas? The answer is very simple. Lighters can be considered as small grinding wheels. In fact, the image showing the collected cooled sparks of lighters is shown in Fig. 7. As one can see, the shape and size distribution of this spherical particles is comparable to those, attributed to micrometeorites.

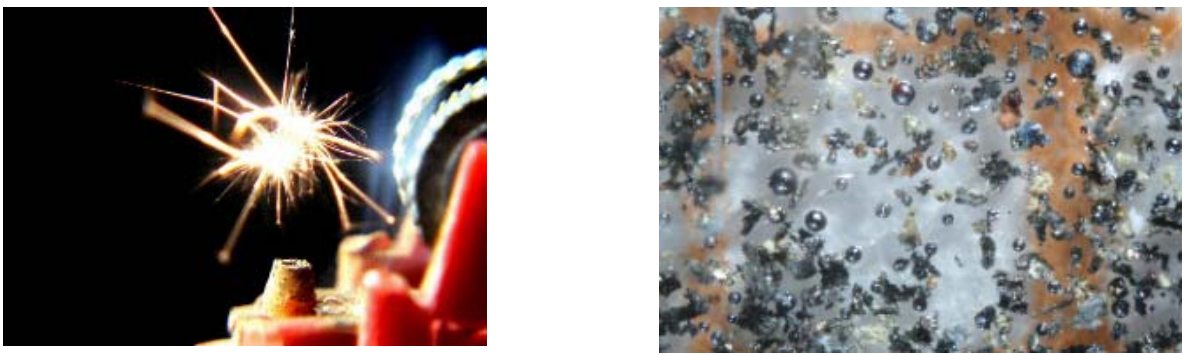

Fig. 7. Sparks of the lighter (a) and image of the sample resultant from several acts of the ligther use (b). Grid unit corresponds to 1 mm

As the general conclusion of the present study, one can clam that the most of spherical particles, that were considered as micrometeorites, have the terrestrial origin and are connected to the human activity. To make grounded conclusions about the cosmic origin of each individual spherical particle, one need to perform its complicated analysis.

The author want to thank Dr. Victor Erokhin for the useful discussion and the help in the manuscript preparation. I want also to thank Prof. Simone Craviotto and Mr. Giuseppe Ricciardelli for useful discussions on the materials.